\documentclass[aps,prl,twocolumn,tightenlines,showpacs,nofootinbib]{revtex4}
\usepackage{bm}
\usepackage{graphics}
\usepackage{epsfig}
\usepackage{graphicx}
\usepackage{amsmath}
\usepackage{amsfonts}
\usepackage{amssymb}
\begin{document}

\title{Thermodynamics of strongly frustrated magnet in a field: Ising antiferromagnet on triangular Husimi lattice}
\author{P. N. Timonin}
\email{timonin@aaanet.ru}
\affiliation{Physics Research Institute at
Southern Federal University, 344090 Rostov-on-Don, Russia}

\date{\today}

\begin{abstract}
Some strongly frustrated magnets such as the "spin-ice" compounds fail to produce any magnetic order at finite temperatures even in the presence of magnetic field. Still they have very unusual low-T thermodynamic properties related to the field-induced ground state transitions. Here we show that general qualitative picture of such peculiar thermodynamics can be obtained in the antiferromagnetic Ising model on the triangular Husimi lattice. The analytical results for this model show magnetic plateaus, entropy spikes, crossing points and peculiarities in magnetic susceptibility and specific heat behavior reflecting the existence of ground state transitions. These signatures of strong frustration may help in search of new frustrated magnets and in the interpretation of experimental data.
\end{abstract}

\pacs{75.10.-b, 75.40.Cx}

\maketitle

The magnetic ions in some crystal lattices do not have ordered magnetic phases at finite temperatures in spite of magnetic interactions between them. The origin of this phenomenon lies in the specific topological properties of such lattices which forbid the existence of unique ground state configuration of local spins. Instead the spin systems on these "geometrically frustrated" lattices have numerous ground states the number of which grows exponentially with system size \cite{1, 2, 3}. Some symmetry breaking perturbations (magnetic field, anisotropic stress etc.) may lift this degeneracy and give rise to some exotic ordered or partially ordered phases \cite{4, 5}. 

Yet in some strongly frustrated magnets such perturbations fail to produce any magnetic order at finite $T$. The notorious example is the "spin-ice" compounds with pyrochlore lattices consisting of the corner-shearing tetrahedra \cite{6}. Such magnetic systems may only have zero-temperature transitions under magnetic field variations between some degenerate ground states. Some signatures of these transitions can be found in the low-temperature thermodynamics. Thus magnetic plateaus, entropy spikes and crossing points shows up in the model of spin-ice compounds \cite{7}. Yet the theoretical description of such effects on realistic 3d lattices needs considerable efforts even in cases of special field orientations where the model can be mapped to the frustrated Ising antiferromagnet \cite{7}. 

So it may be useful to consider simplified models to gain the general insight into the origin and overall picture of the peculiar low-temperature thermodynamics of strongly frustrated magnets. Here we show that the model of frustrated Ising antiferromagnet on triangular Husimi lattice can qualitatively reproduce all peculiarities of spin-ice model in [111] field  \cite{7} which are related to the field-induced ground state phase transitions. This is quite simple model allowing for the analytical description of all thermodynamic parameters thus giving the consistent picture of the frustration-induced anomalies. So these results may help in the interpretation of experimental data and in search for new strongly frustrated magnets.

\begin{figure}[htp]
\centering
\includegraphics[height=3cm]{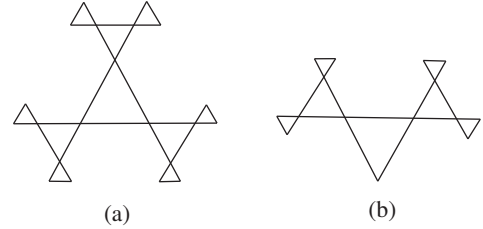}
\caption{Fragments of Husimi lattice (a) and Husimi tree(b). Infinite species are obtaned by sequential addition of triangles to outer sites.}
\label{Fig1}
\end{figure}

\begin{figure}[htp]
\centering
\includegraphics[height=3cm]{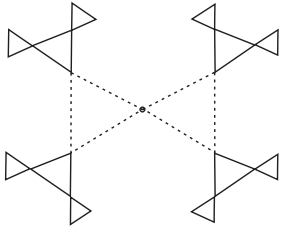}
\caption{Construction of Husimi lattice from 4 Husimi trees.}\label{Fig2}
\end{figure}

Fragment of infinite triangle Husimi lattice composed of Husimi trees is shown in Fig. \ref{Fig1}. We consider Ising antiferromagnet on this lattice. Its partition function can be obtained through consideration of the recursion relations for partial partition function of Husimi tree summed over all spins except the root one 
\begin{equation}
Z_{n + 1} \left( s \right) = Tr\left\{ {e^{ \left[ {\left( {h - Ks} \right)\left( {s_1  + s_2 } \right) - Ks_1 s_2 } \right]}
Z_n \left( {s_1 } \right)Z_n \left( {s_2 } \right)} \right\}.
\label{eq:1}
\end{equation}
Here $Tr$ denotes the summation over spins $s_1$ and$ s_2$, $K = J/T,{\rm{   }}h = H/T$. Defining the effective field
\[
f_n  = \frac{1}{2}\ln \frac{{Z_n \left( 1 \right)}}{{Z_n \left( { - 1} \right)}}
\]
we get for it from Eq. \ref{eq:1}
\begin{equation}
f_{n + 1}  = \frac{1}{2}\ln \frac{{\cosh 2\left( {f_n  + h - K} \right) + e^{2K} }}{{\cosh 2\left( {f_n  + h + K} \right) + e^{2K} }}
\label{eq:2}
\end{equation}
The stationary point of Eq. \ref{eq:2} $f_\infty   = f\left( {K,h} \right)$
defines all thermodynamics of the model. Indeed, with  $f_\infty   = f\left( {K,h} \right)$ we can obtain the free energy per spin $F\left( {T,H} \right)$
as follows. Let us consider 12 Husimi trees and divide them in 4 triples. Adding to each triple 3 bonds we can get 4 Husimi lattices. Otherwise we can divide 12 trees in 3 quadruples and then adding to each quadruple 1 site and 6 bonds we get 3 Husimi lattices , cf. Fig. \ref{Fig2}. 
In the infinite lattice limit the free energies of all Husimi lattices in both constructions will tend to the similar extensive values with the only difference that in the latter construction we have 3 more sites than in the former one. Thus for the free energy per spin $F\left( {T,H} \right)$ we get
\[
\begin{array}{l}
 3F =  - 3T\mathop {\lim }\limits_{n \to \infty } \left\{ {\ln U_n } \right\} +4T\mathop {\lim }\limits_{n \to \infty } \left\{ {\ln V_n } \right\}
\\ 
U_n=Tr\prod\limits_{i = 0}^4 {Z_n \left( {s_i } \right)} e^{\left( {h - Ks_0 } \right)\left( {s_1  + s_2  + s_3  + s_4 } \right) + hs_0  - K\left( {s_1 s_2  + s_3 s_4 } \right)}
\\
V_n = Tr\prod\limits_{i = 1}^3 {Z_n \left( {s_i } \right)} e^{h\left( {s_1  + s_2  + s_3 } \right) - K\left( {s_1 s_2  + s_3 s_1  + s_3 s_2 } \right)} 
\end{array}
\]

Performing here the summation over spin values and using the stationary point equation for $f_\infty   = f\left( {K,h} \right)$
\begin{equation}
f = \frac{1}{2}\ln \frac{{\cosh 2\left( {f + h - K} \right) + e^{2K} }}{{\cosh 2\left( {f + h + K} \right) + e^{2K} }}
\label{eq:3}
\end{equation}
and definition of $f$ we finally get
\begin{eqnarray}
3F/T = 2K - \ln 2 - 2f+\ln \cosh \left( {h + 2f} \right)- 
\nonumber
\\
2\ln \left[ {\cosh 2\left( {h + f + K} \right) + e^{2K} } \right] 
\label{eq:4}
\end{eqnarray}
Differentiating $F$ with respect to $H$ and $T$ we can get magnetization, magnetic susceptibility, entropy and specific heat of the model. In particular, we get the simple expression for magnetization
\begin{equation}
m =  - \frac{{\partial F}}{{\partial H}} = \tanh \left( {2f + h} \right) = \frac{{D^2  - 1}}{{D^2  + 1}},{\rm{  }}D = \exp \left( {2f + h} \right).
\label{eq:5}
\end{equation}
Introduction of new variable $D$ instead of $f$ is useful because Eq. 3 becomes with it just simple cubic equation
\begin{equation}
yzD^3  + \left( {2y - y^{ - 1} z^2 } \right)D^2  + \left( {y^{ - 1} z^{ - 1}  - 2yz} \right)D - y = 0
\label{eq:6}
\end{equation}
Here $y \equiv \exp{2K}$, $z \equiv \exp{h}$. Note that as $m>0$ at $H >0$ physically relevant solution for $D$ should be greater than 1 for positive fields which we consider further. 
Expressed in these new variables free energy (\ref{eq:4}) has the form
\begin{equation}
3F/T = 3\ln (zy) + \ln \left( {1 + D^2 } \right) - 2\ln (z^2 y^2 D^2  + 2zy^2 D + 1)
\label{eq:7}
\end{equation}
In the low temperature region Eqs. (6), (7) can be further simplified. First, when field $H >0$ is also small, $\max \left( {H,T} \right) \ll J$, ($y \gg z > 1$) we have
\begin{eqnarray}
zD^3  + 2D^2  - 2zD - 1 = 0
\label{eq:8}
\\
3F/T =  - 2K + 3\ln (z) + \ln \left( {1 + D^2 } \right) -
\nonumber
\\
 2\ln \left( {z^2 D^2  + 2zD} \right).
\label{eq:9}
\end{eqnarray}
Thus $D$ and $F/T$ in this region depends only on $z$ and this results in very unusual behavior of thermodynamic variables when $H,T \to 0$. Thus magnetization (\ref{eq:5}), entropy

\begin{equation}
S =  - \frac{{\partial F}}{{\partial T}} = \frac{1}{3}\ln \frac{{\left( {1 + 2zD} \right)^2 }}{{D^2 \left( {1 + D^2 } \right)}} - \frac{{2Dz\ln z}}{{\left( {2 + zD} \right)\left( {1 + D^2 } \right)}},	
\label{eq:10}
\end{equation}
reduced magnetic susceptibility
\begin{equation}
\chi ' \equiv T\chi  = T\frac{{\partial m}}{{\partial H}} = \frac{{4D\left( {2D^2  - 1} \right)}}{{z\left( {3D^2  - 2} \right) + 4D}}
\label{eq:11}
\end{equation}
and specific heat 
\begin{equation}
C = T\frac{{\partial S}}{{\partial T}} = \left( {\ln z} \right)^2 \chi '
\label{eq:12}
\end{equation}
are the functions of the ratio $H/T$ only. So they are constant along the lines $H/T=const.$ when $H,T \to 0$.

\begin{figure}[htp]
\centering
\includegraphics[height=8cm]{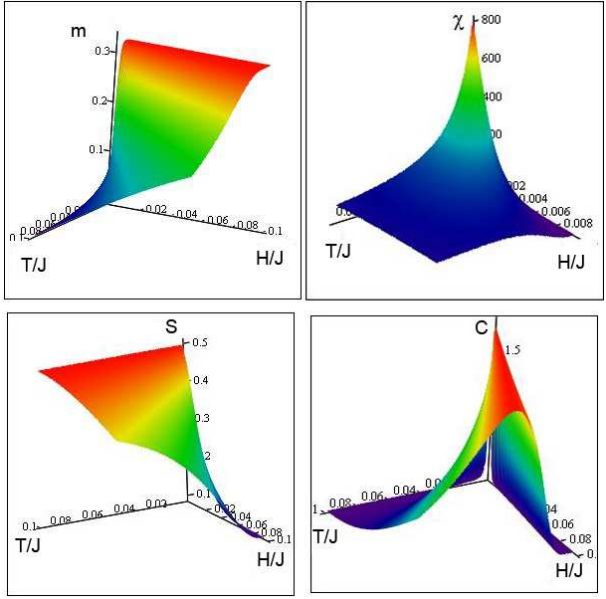}
\caption{(color online) Field and temperature dependences of thermodynamic parameters at $H,T \ll J$.}
\label{Fig3}
\end{figure}

The physical solution ($D>1$) to Eq. \ref{eq:8} (which is also the only stable one with $\chi >0$) has the form
\[
D\left( z \right) = \frac{2}{{3z}}\left\{ {{\mathop{Re}\nolimits} \left[ { - \frac{9}{2}z^2  - 8 + i\frac{3}{2}zR(z)} \right]^{\frac{1}{3}}  - 1}\right\}
\]
\[
R(z) \equiv \sqrt {3\left( {32z^4  + 61z^2  + 32} \right)} 
\]
Here the power $\frac{1}{3}$ means the main branch of the power function. This $D(z)$ grows monotonously from $D(1)=1$ to $D\left( \infty  \right) = \sqrt 2$. So at $H=0$ we have
\[
m=0,\qquad\chi = 4/5T,\qquad S=\frac{1}{3}\ln\frac{9}{2}\approx 0.501,\qquad C = 0 
\]
while at $T=0$
\[
m=1/3,\qquad\chi = 0,\qquad S=\frac{1}{3}\ln\frac{4}{3}\approx 0.096,\qquad C = 0.
\]
In general these thermodynamic variables has different limiting values at $H,T \to 0$ depending on the way to this point in the $H, T$ plane. The general picture of their behavior near $H=T=0$ is shown in Fig. \ref{Fig3}.

\begin{figure}[htp]
\centering
\includegraphics[height=8cm]{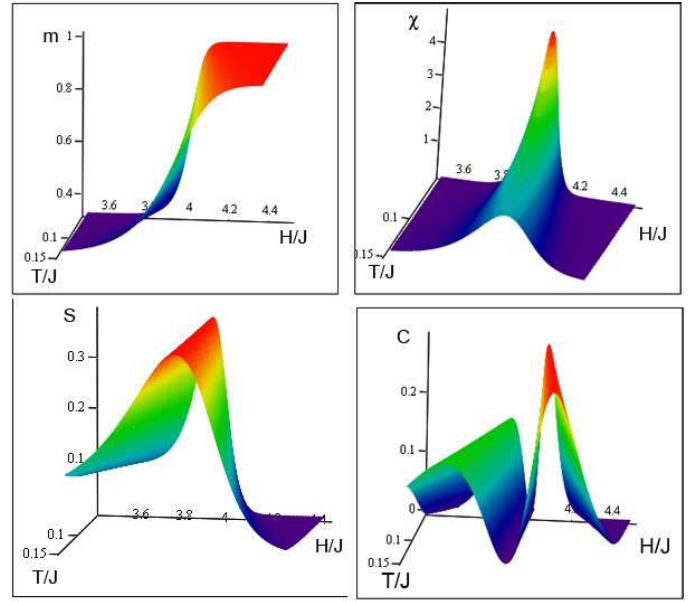}
\caption{(color online) Field and temperature dependences of thermodynamic parameters at $T \ll J$, $H \approx 4J$.}
\label{Fig4}
\end{figure}

\begin{figure}[htp]
\centering
\includegraphics[height=9cm]{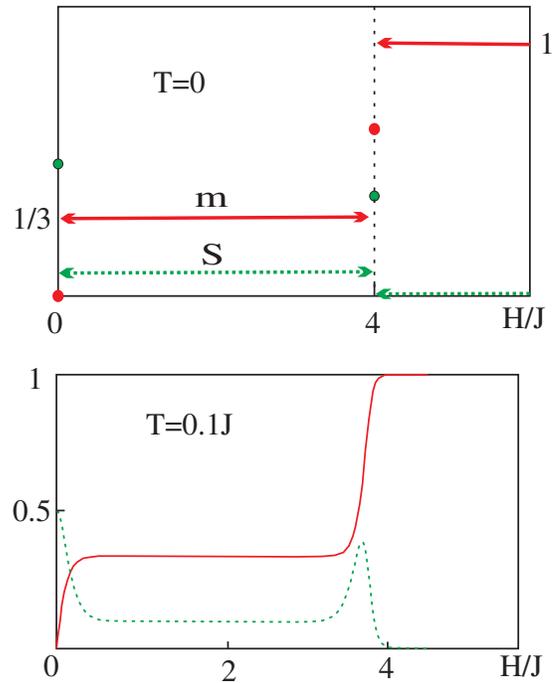}
\caption{(color online) Field dependences of magnetization (solid lines) and entropy (dotted lines) at $T=0$ and $T=0.1J$.}
\label{Fig5}
\end{figure}

Similar scaling features can be found in strong fields $H > 2J \gg T$
($z \gg y \gg 1$). Here we have 
\begin{eqnarray*}
D = \frac{1}{2}\left( {x + \sqrt {x^2  + 8} } \right), \qquad  x \equiv zy^{ - 2}  = \exp \left( {h - 4K} \right)
\\
3F/T =  - 2K + \ln x + \ln \left( {1 + D^2 } \right) - 4\ln D
\end{eqnarray*}
So $D$ and $F/T$ in this region also depends only on one variable $x$ (or $\left( {H - 4J} \right)/T$). Accordingly, this scaling holds for the thermodynamic quantities
\begin{eqnarray*}
S = \frac{4}{3}\ln D - \frac{1}{3}\ln \left( {1 + D^2 } \right) - \frac{2}{3}\frac{{D^2  - 2}}{{D^2  + 1}}\ln x,
\\
\chi ' = \frac{{4xD^2 }}{{\sqrt {x^2  + 8} \left( {D^2  + 1} \right)^2 }}, 
\\
C = \left( {\ln x} \right)^2 \chi '
\end{eqnarray*}
Their behavior near $H=4J$ is shown in Fig. \ref{Fig4}.
At $T = 0$ we get
\[
H < 4J,\qquad m=\frac{1}{3},\qquad S=\frac{1}{3}\ln\frac{4}{3},\qquad\chi'=C=0,
\]
$H = 4J$,  $m=\frac{3}{5}$, $S=\frac{1}{3}\ln\frac{16}{5}\approx 0.388$,
 $\chi'=\frac{16}{75}$, $C = 0$
\[
H > 4J,\qquad m=1,\qquad S=\chi'=C=0
\]
Note that the results for $T = 0, H<4J$ coincide with those for $T = 0, H \ll J $. So at $0<H<4J$ we have a plateau in the field dependence of magnetization ($m = \frac{1}{3}$) and other thermodynamic parameters and at $H>4J$ there is another plateau. They are shown in Fig. \ref{Fig5} for $m$ and $S$.

Thus at $T=0$ we have two field-induced first-order transitions at $H=0$ and at $H=4J$. The nature of these transitions is quite apparent. In zero field we have highly degenerate ground states having two parallel spins and the anti-parallel one in each triangle. Their number can be computed using Pauling-Anderson-type estimates \cite{8}. First we consider the spins as belonging to $N/3$ independent triangles each having 6 ground state configurations. Each site in such configurations takes the values +1 and -1 with probability $\frac{1}{2}$. Hence the probability $p$ that the bond triangles connecting the independent ones have also the lowest energy configurations is $p=6\left({\frac{1}{2}}\right)^3$. Thus the number of ground states is
$\Gamma  = \left( {6p} \right)^{N/3}  = \left( {\frac{9}{2}} \right)^{N/3} $
giving the quoted above result for the zero-field entropy $S = N^{ - 1} \ln \Gamma $. Thus Pauling-Anderson entropy estimate \cite{8} neglecting the correlations between spin ordering in the next nearest triangles is exact for zero-field ground states on the Husimi lattice.

The arbitrary small field lifts partially this degeneracy. Here we can also apply the Pauling-Anderson approach. Now only three triangle configurations (permutations of (1, 1,-1)) give the lowest energy. In such configurations each spin is 1 with probability $2/3$ and -1 with probability $1/3$. So the probability $p$ that the bond triangles connecting the independent ones also have ground state configurations is now $p = 3\frac{1}{3}\left( {\frac{2}{3}} \right)^2  = \frac{4}{9}$ and  the number of ground states is
$\Gamma  = \left( {3p} \right)^{N/3}  = \left( {\frac{4}{3}} \right)^{N/3} $
in accordance with above $S$ for $0<H<4J$. So here we have another phase with lower $S$ and $m = \frac{1}{3}$.  

One more phase (ferromagnetic) appears at $H>4J$ with $S=0$ and $m=1$ while at $H=4J$ special ground states exist. At this point the energy of ferromagnetic configuration of two triangles with one common site (dashed lines in Fig. \ref{Fig2}) does not actually depend on the direction of the central spin. The magnetization of these states $m = \frac{3}{5}$ shows that they have the fraction $1/5$ of spins pointed opposite to the field direction. So we may conclude that here the ground state configurations can be obtained by the divisions of lattice into couples of triangles with common site having -1 spin and other four sites with +1 spins. Then at $H=4J$ $\Gamma  = e^{NS}  = \left( {\frac{{16}}{5}} \right)^{N/3} $ can be identified with the number of divisions of Husimi lattice into a couples of connected triangles. Yet here we can not prove this correspondence. We may only observe that the huge degeneracy of ground states at $H=4J$ give rise to the entropy spike which transforms into sharp maximum at finite $T$, cf. Fig.\ref{Fig5}, quite similar to that found in the model for spin-ice pyrochlores \cite{7}. 

In general, the low-$T$ features of thermodynamic parameters in a field stem from the existence of $T=0$ phase transitions. Thus all magnetization curves for different low-$T$ values will have $m = \frac{5}{6}$ at $H=4J$ so this is the crossing point for these curves. This crossing effect has been found in the spin-ice model \cite{7}. But the present model shows that it should also exist for all dimensionless parameters $S, C$ and $\chi '$: at all small $T \ll J$
they acquire their ground state values at $H=4J$. Note also that the divergences in temperature dependencies of susceptibility $\chi  \sim T^{ - 1} $
 appear only at the field transitions' points $H=0$ and $H=4J$, otherwise $\chi  \to 0$ at $T \to 0$ coming over a broad maximum. Meanwhile the magnetic specific heat $C$ is zero at $H=0$ and $H=4J$ which give rise to its double-peaked field dependencies around these points (cf. Figs. \ref{Fig3}, \ref{Fig4}). Along with the entropy spikes and magnetization plateaus this may serve as indication of the ground state phase transitions in which ground state degeneracy is lifted by an external field. 

To conclude we may state that present model of Ising antiferromagnet on Husimi lattice can give qualitatively adequate picture of the specific low-temperature thermodynamics which may exist in real 3d frustrated magnets when their spin system allows for the Ising-type description. Apparantly the model with effective dimension $d = \infty$ may give only approximate information on the structure of the ground state configurations in real crystals and may grossly overestimate their degeneracy as compared with the lattices in 3d Eucledean space. This fault can be partially remedied by considering the generalizations of Husimi lattice where the basic building blocks are the fragments of 3d lattices instead of triangles. Yet failing to reproduce quantitatively the values of thermodynamic variables in real lattices it may correctly describe the thermodynamic anomalies near field-induced phase transitions because of their first-order nature allowing for the mean-field description. Also the consideration of anisotropic Heisenberg model on such lattice may provide useful qualitative information on the properties of real geometrically frustrated magnets.


%
%

\begin{thebibliography}{8}
\expandafter\ifx\csname natexlab\endcsname\relax\def\natexlab#1{#1}\fi
\expandafter\ifx\csname bibnamefont\endcsname\relax
  \def\bibnamefont#1{#1}\fi
\expandafter\ifx\csname bibfnamefont\endcsname\relax
  \def\bibfnamefont#1{#1}\fi
\expandafter\ifx\csname citenamefont\endcsname\relax
  \def\citenamefont#1{#1}\fi
\expandafter\ifx\csname url\endcsname\relax
  \def\url#1{\texttt{#1}}\fi
\expandafter\ifx\csname urlprefix\endcsname\relax\def\urlprefix{URL }\fi
\providecommand{\bibinfo}[2]{#2}
\providecommand{\eprint}[2][]{\url{#2}}


\bibitem[1]{1}
\bibinfo{author}{\bibfnamefont{R.~M.~F.} \bibnamefont{Houtappel}},
  \bibinfo{journal}{ Physca} \textbf{\bibinfo{volume}{16}},
  \bibinfo{pages}{ 425} (\bibinfo{year}{1950}).

\bibitem[2]{2}
\bibinfo{author}{\bibfnamefont{G.~H.} \bibnamefont{ Wannier}},
\bibinfo{journal}{Phys.\ Rev.} \textbf{\bibinfo{volume}{79}},
\bibinfo{pages}{ 357} (\bibinfo{year}{1950}).


\bibitem[3]{3}
\bibinfo{author}{\bibfnamefont{ R.}\bibnamefont{ Liebmann}},
\bibinfo{journal}{Lecture Notes in Physics}, \textbf{\bibinfo{volume}{251}},
\bibinfo{editor}{Springer},(\bibinfo{year}{1986}).
  

\bibitem[4]{4}
\bibinfo{author}{\bibfnamefont{R.}~\bibnamefont{Moessner}},
  \bibinfo{journal}{Can.\ J.\ Phys.} \textbf{\bibinfo{volume}{79}},
  \bibinfo{pages}{1283} (\bibinfo{year}{2001}).


\bibitem[5]{5}
\bibinfo{author}{\bibfnamefont{R.}~\bibnamefont{Moessner}},
  \bibinfo{journal}{J.\ Phys.\ Conf.\ ser.} \textbf{\bibinfo{volume}{145}},
  \bibinfo{pages}{012001} (\bibinfo{year}{2009}).

\bibitem[6]{6}
\bibinfo{author}{\bibfnamefont{J.~S.}~\bibnamefont{Gardner}},
\bibinfo{author}{\bibfnamefont{M.~J.~P.}~\bibnamefont{Gingras}}
 \bibnamefont{and}
\bibinfo{author}{\bibfnamefont{J.~E.}~\bibnamefont{Greedan}},
  \bibinfo{journal}{Rev.\ Mod.\ Phys.} \textbf{\bibinfo{volume}{82}},
  \bibinfo{pages}{53} (\bibinfo{year}{2008}).

\bibitem[7]{7}
\bibinfo{author}{\bibfnamefont{S.~V.} \bibnamefont{ Isakov }},
\bibinfo{author}{\bibfnamefont{K.~S.} \bibnamefont{ Raman }},
\bibinfo{author}{\bibfnamefont{R.} \bibnamefont{ Moessner }}
 \bibnamefont{and}
\bibinfo{author}{\bibfnamefont{ S.~L.}~\bibnamefont{ Sondhi }},
\bibinfo{journal}{Phys.\ Rev.\ B} \textbf{\bibinfo{volume}{70}},
\bibinfo{pages}{104418} (\bibinfo{year}{2004}).


\bibitem[8]{8}
\bibinfo{author}{\bibfnamefont{ L.}~\bibnamefont{ Pauling}}
\bibinfo{journal}{ J.\ Am.\ Chem.\ Soc. } \textbf{\bibinfo{volume}{57}},
\bibinfo{pages}{2680} (\bibinfo{year}{1935});
\bibinfo{author}{\bibfnamefont{ P.~W.}~\bibnamefont{ Anderson}},
 \bibinfo{journal}{ Phys.\ Rev.} \textbf{\bibinfo{volume}{102}},
  \bibinfo{pages}{1008} (\bibinfo{year}{1956}) .

\end{thebibliography}
\end{document}